\documentclass[fleqn,twoside]{article}%
\usepackage{amsmath}
\usepackage{amsfonts}
\usepackage{amssymb}
\usepackage{graphicx}%
\begin{document}

\title{Revisiting conserved currents in $F(R)$ theory of gravity via Noether symmetry}

\author{Nayem Sk.$^\dag$ and Abhik Kumar Sanyal$^{\S}$}
\maketitle

\noindent

\begin{center}
$\dag$Dept. of Physics, University of Kalyani, Nadia, India - 741235\\
$\S$Dept. of Physics, Jangipur College, Murshidabad, India - 742213\\

\end{center}

\footnotetext[1]{
Electronic address:\\

\noindent $\dag$nayemsk1981@gmail.com\\
\noindent $^{\S}$sanyal\_ ak@yahoo.com\\}

Noether symmetry of $F(R)$ theory of gravity in vacuum and in the presence of pressureless dust yields $F(R) \propto R^{\frac{3}{2}}$ along with the conserved current $\frac{d}{dt}(a\sqrt R)$ in Robertson-Walker metric and nothing else. Still some authors recently claimed to have obtained four conserved currents setting $F(R) \propto R^{\frac{3}{2}}$ a-priori, taking time translation along with a gauge term. We show that the first one of these does not satisfy the field equations and the second one is the Hamiltonian which is constrained to vanish in gravity and thus a part and parcel of the field equations. We also show that the other two conserved currents, which do not contain time translation are the same in disguise and identical to the one mentioned above. Thus the claim is wrong.\\


$F(R)$ theory of gravity has drawn lot of attention to the cosmologists in recent years, since it elegantly solves the cosmic puzzle in connection with dark energy, unifies early inflation with late time cosmic acceleration, admits Newtonian limit and passes the solar test singlehandedly [see \cite{1} for recent and thorough reviews]. However, in order to select a particular form of $F(R)$ out of indefinitely large number of curvature invariant terms, Noether symmetry was invoked as a selection rule by several authors \cite{2} - \cite{9}. Results in a nutshell are (i) Noether symmetry does not admit anything other than $F(R) \propto R^{\frac{3}{2}}$, along with a conserved current $\frac{d}{dt}(a\sqrt R)$ in Robertson-Walker metric; (ii) Explicit solution of the field equations shows that it is well behaved in the early Universe, nevertheless it suffers from the same disease as $R^{-n},~n > 0$ term in the late Universe in the presence of radiation and a pressureless dust. In the radiation era the scale factor tracks as $a \propto t^{\frac{3}{4}}$ instead of the standard Friedmann-type ($a \propto \sqrt t$) and the early deceleration in the matter dominated era evolves like $a \propto \sqrt t$ instead of the standard Friedmann type ($a \propto t^{\frac{2}{3}}$). These results create problem in explaining the structure formation; (iii) Situation is improved taking a linear term in addition, but paying a price of giving up Noether symmetry; (iv) Noether symmetry is not admissible if the configuration space is enlarged adding a scalar field or taking anisotropic models into account \cite{9}.\\

Such an apparently closed chapter of $F(R) \propto R^{\frac{3}{2}}$ has been reopened by some authors \cite{10} - \cite{12} claiming to produce some new conserved currents in the name of Noether gauge symmetry. Particularly, it was claimed by Hussain et-al \cite{10} that Noether gauge symmetry admits $F(R) \propto R^n$, where $n$ is arbitrary. Jamil et-al \cite{11} on the other hand found $F(R) \propto R^2$ and $V(\phi) \propto \phi^{-4}$, considering Noether gauge symmetry with Tachyon. Ridiculously, in both of these published work the authors claimed their result to be an outcome of gauge Noether symmetry, although they either set the gauge term to vanish a-priori \cite{10} or forced the gauge term to vanish by suitably manipulating the variables \cite{11}. Thus only additional issue considered in their work is the time translation. It is well-known that the Hamiltonian is the generator of time translation and it is constrained to vanish in the theory of gravity. Thus, it is the $(^0_0)$ equation of Einstein and is a part and parcel of the Field equations. This is the reason why earlier authors while invoking Noether symmetry in the theory of gravity never took time translation into account. The claim \cite{10} has been revisited by the present authors \cite{13} taking both vanishing and non-vanishing gauge into account. The result is that, for $F(R) \propto R^n$, only $n = \frac{3}{2}$ is admissible. Although $n = 2$  satisfies all the Noether equations but then, it does not satisfy the field equations, while for other values of $n$, Noether equations are not satisfied. Thus, the claim that arbitrary power of $R$ generates Noether symmetry is wrong. The claim \cite{11} has also been reviewed by the present authors \cite{14} and it was shown that $R^2$ does not satisfy the Tachyonic field equations and thus Noether symmetry remains obscure. This proves our earlier claim \cite{9} that Noether symmetry remains obscure if the configuration space is enlarged by adding a scalar field in the $F(R)$ theory of gravity.\\

More recently Shamir et-al \cite{12} has claimed that Noether symmetry of $F(R) = R^{\frac{3}{2}}$ admits four different generators corresponding to which four different conserved currents exist in the presence of non-zero gauge. Here we review the work and prove that such claim is totally wrong. In the following, we first write down the expressions for the vector field and the conserved current in the presence of a gauge term, which has been explicitly calculated earlier \cite{13}. Next, we write down the field equations and gauge Noether equations for $F(R)$ theory of gravity in the presence of a matter field. We then review the work of Shamir et-al and prescribe a short hand procedure to find explicit solution for this particular form ($R^{\frac{3}{2}}$) of the curvature invariant term and finally conclude.\\

Let us now introduce Noether symmetry generator with gauge and the conserved current. Among all the dynamical symmetries, transformations that map solutions of the equations of motion into solutions, one can single out Noether symmetries as the continuous transformations that leave the action invariant - except for boundary terms. In formal language, Noether symmetry for a point Lagrangian $L(q_i, \dot q_i, t)$ states that, For any regular system if there exists a vector field $X^{(1)}$, such that,

\begin{equation} \left(\pounds _{X^{(1)}} + \frac{d\eta}{dt}\right)L = \left(X^{(1)} + \frac{d\eta}{dt}\right) L = \frac{d B}{d t},\end{equation}
in the presence of a gauge function $B(q_i, t)$, then there exists a conserved current,

\begin{equation} I = \sum_{i}(\alpha_i - \eta \dot q_{i} )\frac{\partial L(q_{i}, \dot q_{i}, t)}{\partial \dot q_{i}} + \eta L(q_{i}, \dot q_{i}, t) - B(q_{i}, t) = \sum_{i}\alpha_i p_{i} - B(q_i, t) - \eta(q_i, t)H(q_i, p_i, t),\end{equation}
where, $X^{(1)}$ is the first prolongation of $X$ given by,

\begin{equation} X^{(1)} = X + \sum_{i}\Big[(\dot\alpha_{i} - \dot \eta\dot q_{i})\frac{\partial}{\partial\dot q_{i}}\Big],\;\;\;with\;\;\;
X = \eta\frac{\partial}{\partial t} +\sum_{i} \alpha_i\frac{\partial}{\partial  q_{i}} ,\end{equation}
with, $\alpha_i = \alpha_i(q_i, t), ~\eta = \eta(q_i, t)$. Having obtained the general form of Noether conserved current with gauge, it has been shown explicitly that for a time independent Lagrangian as in the case of gravity, the gauge term has to be time independent \cite{13}. Further, for the gravitational Lagrangian under consideration since Hamiltonian is constrained to vanish, so it is treated as a part and parcel of the field equations (the $(^0_0)$ component of Einstein's equations) rather than a separate conserved quantity. Hence time translation is overall unnecessary. Thus Noether current reduces in this case to

\begin{equation} I =  \sum_{i}\alpha_i p_{i} - B(q_i).\end{equation}
Although we have proved that time translation is unnecessary to find Noether currents in the case of gravity, still in our analysis we keep time translation to show its role explicitly.\\
Let us now take up the action corresponding to $F(R)$ theory of gravity in the presence of matter, write down the point Lagrangian, the field equations and Noether gauge equations. Taking Robertson-Walker line element

\begin{equation} ds^2 = -dt^2+a^2\left[\frac{dr^2}{1-kr^2} + r^2 d\theta^2 + r^2\sin^2\theta d\phi^2\right],\end{equation}
into account, the following action

\begin{equation} A = \int d^4 x \sqrt{-g} ~[\gamma F(R)+L_m ],\end{equation}
leads to a point Lagrangian ($L_m$ being the matter Lagrangian)

\begin{equation} L = \left[6a \dot a^2 F'  + 6a^2\dot a\dot R F'' - a^3(F' R - F) - 6kaF'\right] + \frac{\rho_0}{\gamma} a^{-3w} ,\end{equation}
treating

\begin{equation} R + 6\left(\frac{\ddot a}{a} + \frac { \dot a^2 + k}{a^2}\right) = 0,\end{equation}
as a constraint of the theory and spanning the Lagrangian by a set of configuration space variable $(a, R, \dot a, \dot R)$. In the above, $\gamma$ is the coupling constant and $\rho_0$ stands for the amount of energy density available in the present Universe, while $w$ is the state parameter. The field equations are,

\begin{equation} \left(2\frac{\ddot a}{a}+\frac{\dot a^2}{a^2} + \frac{k}{a^2}\right)F' + \left(\ddot R + 2 \frac{\dot a}{a}\dot R\right)F'' + \dot R^2 F''' +\frac{1}{2}(F'R - F) = - \frac{w\rho_0}{2\gamma} a^{-3(w+1)} = -\frac{p}{2\gamma},\end{equation}

\begin{equation} 3\left(\frac{\dot a^2}{a^2} + \frac{k}{a^2}\right)F' + 3\frac{\dot a}{a}\dot R F'' + \frac{1}{2}(F' R - F) = \frac{\rho_0}{2\gamma} a^{-3(w+1)} = \frac{\rho}{2\gamma}.\end{equation}
The energy density and the pressure of the fluid have been represented by $\rho$ and $p$ in the above set of field equations. The Noether gauge equation (1) is,

\[\alpha\Big[6 \dot a^2 F' + 12 a \dot{a}\dot R F'' - 3a^2 (F' R - F) - 6k F' - 3\frac{\rho_0}{\gamma} w a^{-(3w+1)}\Big] + \beta\Big[6 a \dot a^2 F'' + 6a^2 \dot a \dot R F''' - a^3 R F'' - 6kaF''\Big] \]
\[ +\Big[\alpha_{,t} + (\alpha_{,a} - \eta_{,t})\dot a + \alpha'\dot R -\dot a^2 \eta_{,a} - \dot a\dot R \eta'\Big](12a\dot a F'+ 6a^2 \dot R F'')
+\Big[\beta_{,t} + \beta_{,a}\dot a + (\beta' - \eta_{,t})\dot R -\dot R^2 \eta' - \dot a\dot R \eta_{,a} \Big](6 a^2\dot a F'')\]
\[+\left[\eta_{,t} + \eta_{,a}\dot a + \eta'\dot R\right]\Big[6 a \dot a^2 F' + 6 a^2\dot a\dot R F'' - a^3(F'R - F) - 6kaF' + \frac{\rho_0}{\gamma} a^{-3w}\Big] = B_{,t} + B_{,a}\dot a + B'\dot R.\]
Equating the coefficients of $\dot a^3,\;\dot a^2 \dot R,\;a\dot R^2,\;\dot a^2,\;\dot a,\;\dot R^2,\;\dot R,\;\dot a\dot R,\;$and the term independent of time derivative to zero respectively as usual, we obtain the following set of Noether equations

\begin{equation} F'\eta_{,a} = 0 \end{equation}
\begin{equation} F'\eta' - a F'' \eta_{,a} = 0 \end{equation}
\begin{equation} \eta'F'' = 0 \end{equation}
\begin{equation} (\alpha + 2a\alpha_{,a}- a\eta_{,t})F' + a(\beta + a \beta_{,a})F''  = 0 \end{equation}
\begin{equation} 12 a\alpha_{,t}F' + 6a^2 \beta_{,t}F'' - \Big[a^3(F' R - F) + 6k a F' - \frac{\rho_0}{\gamma}a^{-3w}\Big]\eta_{,a}= B_{,a} \end{equation}
\begin{equation} \alpha' F'' = 0  \end{equation}
\begin{equation} 6a^2 \alpha_{,t}F'' - \Big[a^3(F' R - F) - 6k a F' + \frac{\rho_0}{\gamma}a^{-3w}\Big]\eta' =  B'\end{equation}
\begin{equation} 2\alpha'F'+a\Big(2\frac{\alpha}{a}+\alpha_{,a}+\beta'-\eta_{,t}\Big)F''+a \beta F''' = 0  \end{equation}
\begin{equation} -a^2(3\alpha+a\eta_{,t})(F' R - F)- a(a^2R + 6k )\beta  F'' - a\eta_{,t}\Big(6kF'- \frac{\rho_0}{\gamma}a^{-(3w+1)}\Big)-\alpha\Big(6kF'+3\frac{w\rho_0}{\gamma}a^{-(3w+1)}\Big) =  B_{,t}\end{equation}
where, dash($'$) represents derivative with respect to $R$. Although we have shown \cite{13} that the gauge term $B$ has to be time independent for a time independent Lagrangian (as in the present case), still we have kept its time dependence form to show explicitly that indeed the solutions of Noether equations does not allow time dependence of $B$. In view of the fact that $F'' \ne 0$, the equations (11) through (13) and (16) are trivially satisfied under the conditions that $\eta$ is independent of $a$ and $R$ while $\alpha$ is independent of $R$. Hence, the above set of equations (11) through (19) may be expressed in a more comprehensive form as,

\begin{equation} (\alpha + 2a\alpha_{,a}- a\eta_{,t})F' + a(\beta + a \beta_{,a})F''  = 0 \end{equation}
\begin{equation} 12 a\alpha_{,t}F' + 6a^2 \beta_{,t}F'' =  B_{,a} \end{equation}
\begin{equation} 6a^2 \alpha_{,t}F'' =  B'\end{equation}
\begin{equation} \Big(2\frac{\alpha}{a}+\alpha_{,a}+\beta'-\eta_{,t}\Big)F''+ \beta F''' = 0  \end{equation}
\begin{equation} a^2(3\alpha+a\eta_{,t})(F-F' R)- a(a^2R + 6k )\beta  F'' - a\eta_{,t}\Big(6kF'- \frac{\rho_0}{w}a^{-(3w+1)}\Big)-\alpha\Big(6kF'+3\frac{\rho_0}{w}a^{-(3w+1)}\Big) = B_{,t}\end{equation}
Having obtained such compact form of Noether gauge equations (20) through (24), let us review of Shamir et al's work \cite{12}. The authors \cite{12} analyzed the above set of Noether equations (20) through (24) choosing $F(R) \propto R^{3/2}$ a-priori - as it should be, and obtained setting both $k = 0$ and $\rho_0 = 0$ the following solutions,

\begin{equation} \eta = c_1 t+ c_2;\;\;\alpha = \frac{2c_1 a^2 + 3c_3 t + 3c_4}{3a};\;\;\beta = -2R\frac{c_1 a^2 + c_3 t + c_4}{a^2};\;\;and\;\;B = 9c_3 a\sqrt R + c_5,\end{equation}
which, as can be checked easily, indeed satisfy all the Noether equations (20) through (24). For the above set of solutions (25), the general conserved current (4) is

\begin{equation} I_0 = 3c_1\frac{a^2}{\sqrt R} \frac{d}{dt}(a R)+ (9c_3 t + c_4)\frac{d}{dt}(a \sqrt R)- 9c_3 a\sqrt R \end{equation}
in view of the fact that the Hamiltonian $H$ is constrained to vanish. So far so good, but then one has to check if the above conserved current satisfies the field equations for $F(R) = R^{\frac{3}{2}}$, which in vacuum are,

\begin{eqnarray}
2\frac{\ddot a}{a}+\frac{\dot a^2}{a^2}+\frac{1}{2}\left[\frac{\ddot R}{ R}+2\frac{\dot a \dot R}{a R}\right]-\frac{1}{4}\frac{\dot R^2}{R^2}+\frac{R}{6}&=& 0.\\
\frac{\dot a^2}{a^2}+ \frac{\dot a \dot R}{2aR}+\frac{R}{18} &=& 0.
\end{eqnarray}
A little algebra shows that the above conserved current (26) does not satisfy the field equations (27) or (28). It is obvious, since Hamiltonian is the generator of time translation $\frac{\partial}{\partial t}$, so a conserved Hamiltonian implies $c_1 = 0$ and $\eta = c_2 = 1$. Hence, restricting $c_1 = 0$, the above conserved current (26) reduces to

\begin{equation} (9c_3 t + c_4)\frac{d}{dt}(a \sqrt R)- 9c_3 a\sqrt R = I_0\end{equation}
It is possible to get rid of the temporal dependance of the conserved current first by integrating (29)

\begin{equation} a\sqrt R = It + \frac{I c_4 - I_0}{9c_3},\end{equation}
($I$ being the constant of integration) and then differentiating (30) to obtain,

\begin{equation} \frac{d}{dt} (a\sqrt R) = I,\end{equation}
which is the same old conserved current obtained earlier \cite{6, 7, 8, 13}. Thus we observe that unless $\eta = 1$, the conserved current does not satisfy the field equations. Explicit time dependence in $\alpha$ and $\beta$ does not play any role. Despite the fact that all the authors \cite{6, 7, 8, 13} obtained the same form of $F(R) \propto R^{\frac{3}{2}}$ under Noether symmetry of $F(R)$ theory of gravity along with the unique conserved current (31), there is a recent claim that the symmetry admits more number of conserved currents \cite{12}. Such claim rooted from the fact that the authors \cite{12} did not check if the conserved currents satisfy the field equations. In the following we forfeit the claim by analyzing the generators obtained by Shamir et-al \cite{12}. Taking the symmetry generator

\begin{equation} X = \eta \frac{\partial}{\partial t}+\alpha \frac{\partial}{\partial a}+\beta\frac{\partial}{\partial R}, \end{equation}
the authors have chosen four different forms of $\eta, \alpha$ and $\beta$ to obtain four symmetry generators and correspondingly four conserved currents. These are $(i) X_1$, obtained setting $\eta = t, \alpha = \frac{2}{3}a$ and $\beta = -2R$, $(ii) X_2$, obtained setting $\eta = 1, \alpha = \beta = 0$, $(iii) X_3$, obtained setting $\eta = 0, \alpha = \frac{t}{a}$ and $\beta = \frac{-2R t}{a^2}$ and $(iv) X_4$, obtained setting $\eta = 0, \alpha = \frac{1}{a}$ and $\beta = \frac{-2R}{a^2}$. Hence the generators and the corresponding conserved currents are as listed underneath,

\begin{equation} X_ 1 = t \frac{\partial}{\partial t}+\frac{2 a}{3} \frac{\partial}{\partial a}-2R \frac{\partial}{\partial R},\;\;\;I_1 = -3a^2\dot a \sqrt R-3a^3\frac{\dot R}{\sqrt R}+\frac{t}{2}(18 a\dot a^2 \sqrt R+9 a^2\dot a\frac{\dot R}{\sqrt R}+a^3 R^\frac{3}{2})\end{equation}
\begin{equation} X_ 2 =  \frac{\partial}{\partial t},\;\;\;I_2 = 9a\dot a^2 \sqrt R+\frac{9}{2}a^2\dot a\frac{\dot R}{\sqrt R}+\frac{1}{2}a^3 R^{\frac{3}{2}}\end{equation}

\begin{equation} X_ 3 = \frac{t}{a} \frac{\partial}{\partial a}-\frac{2Rt}{a^2} \frac{\partial}{\partial R},\;\;\;I_3 = 9a\sqrt R-t\Big[9\dot a\sqrt R+\frac{9a}{2}\frac{\dot R}{R^{\frac{1}{2}}}\Big]\end{equation}
\begin{equation} X_ 4 = \frac{1}{a} \frac{\partial}{\partial a}-\frac{2R}{a^2} \frac{\partial}{\partial R},\;\;\;I_4 = -9\dot a\sqrt R-\frac{9 a}{2}\frac{\dot R}{\sqrt R}\end{equation}
It is not difficult to express the first two conserved currents as,

\begin{equation} I_1 = -3a^2\dot a \sqrt R-3a^3\frac{\dot R}{\sqrt R}+18 a^3 \sqrt R t H\end{equation}

\begin{equation} I_2 = 9a^3 \sqrt R  H.\end{equation}
As we have repeatedly mentioned that the Hamiltonian ($H$) is constrained to vanish, thus $I_2$ does not appear at all while $I_1$ reduces to

\begin{equation} I_1 = -3\frac{a^2}{\sqrt R}\frac{d}{dt} (a R)\end{equation}
whose time derivative yields

\begin{equation} \frac{\ddot a}{a}+\frac{2\dot a^2}{a^2}+\frac{7\dot a \dot R}{2aR}+\frac{\ddot R}{R}-\frac{\dot R^2}{2R^2} = 0.\end{equation}
It is quite apparent that equation (40) does not satisfy the field equations (27) and (28). So $X_1$ is not a generator of the system under consideration. Therefore, in gravity one has to set $c_1 = 0$ or rather one should never consider time translation ie., set $\eta = 0$, since as mentioned earlier, Hamiltonian is not only conserved but is constrained to vanish and as a result it is the $(^0_0)$ component of Einstein's field equation. So the author's \cite{12} conserved currents reduce to two, viz., $I_3$ and $I_4$ corresponding to the generators $X_3$ and $X_4$ and it is important to note that these two generators do not include time translation ($\eta = 0$). It is clearly observed that

\begin{equation} - I_4 = I = \frac{d}{dt}(a\sqrt R),\end{equation}
where $I$ is the conserved current already obtained in equation (31) and also by earlier authors \cite{6, 7, 8, 13}. So the only conserved current left is $I_3$ given in equation (35), which may be written as,

\begin{equation} -\frac{I_3}{9t^2} = -\frac{a\sqrt R}{t^2} + \frac{1}{t}\frac{d}{dt} (a\sqrt R) = \frac{d}{dt}\Big(\frac{a\sqrt R}{t}\Big).\end{equation}
Under integration the above equation may be reexpressed as,

\begin{equation}\frac{a\sqrt R}{t} = \frac{I_3}{9t} + I,\end{equation}
where, $I$ is a constant. Substituting $I_3$ from equation (42) in the above, one again arrives at the same old conserved current (41) which is the one already found in (31) and by the earlier authors \cite{6, 7, 8, 13}. Thus, we show that the first conserved current $I_1$ given in (33) containing the time translation in the form $t\frac{\partial}{\partial t}$, i.e., $\eta = t$ does not satisfy the field equations. This is obvious, since the gravitational Hamiltonian is conserved for which it is required to set $\eta = 1$. The second conserved current which again contains only time translation in the form $\frac{\partial}{\partial t}$, i.e., $\eta = 1$, gives the Hamiltonian which is not only conserved but is constrained to vanish. As a result $I_2 = 0$ implying $H = 0$ is just the $(^0_0)$ equation of Einstein's field equation. One is thus left with the conserved currents $I_3$ and $I_4$ given in equations (35) and (36) respectively. Since none of these contains time translation ($\eta = 0$), so these are of interest. We have shown that $I_4$ is actually the same old conserved current found earlier \cite{6, 7, 8, 13} and also in equation (31) here. Apparently though $I_3$ appears to be a new one, nevertheless we have shown that this is again the same conserved current in disguise. Thus the claim to have obtained four different conserved current of $F(R) \propto R^{\frac{3}{2}}$ theory of gravity in the presence of gauge and time translation \cite{12} has been falsified. It may be mentioned that keeping non-zero curvature parameter ($k$) and pressureless dust as the matter source in addition, do not change the result discussed, as has been elaborated by Kaushik et al \cite{8}.\\

In view of above analysis, we learn that one must not consider time translation to find Noether symmetry in cosmology. Further, it is useless to add a gauge term , since it gives nothing new, rather makes things complicated. Here, we just review our earlier work \cite{8} to expatiate the simplest technique that should be adopted in the case of $F(R) \propto R^{\frac{3}{2}}$. For the purpose, let us choose $a^2 = z$. Note that for canonical quantization of higher order theory of gravity, one has to start with the basic variable $h_{ij} = a^2$ \cite{15, 16}. In view of such a change of variable, the action for $R^{\frac{3}{2}}$ term in the presence of a pressureless fluid reads,

\begin{equation} S = \mathcal{B}\int \left[3\sqrt3(\ddot z + 2k)^{\frac{3}{2}} - \rho_{m0} \right]dt - 2 \mathcal{B}\int \left[\sqrt h F_{,R} K \right]d^3 x,\end{equation}
where, $\rho_{m0}$ is the amount of matter density (CDM + Baryonic fluid) presently available in the Universe. Next we define an auxiliary variable as

\begin{equation} Q = \frac{\partial S}{\partial \ddot z} = \frac{9\sqrt 3}{2}\mathcal{B}(\ddot z + 2k)^{\frac{1}{2}} = \frac{9}{2}\mathcal{B} a\sqrt R\end{equation}
in view of which the action may be expressed in the following canonical form

\begin{equation} S = \int\left[-\dot Q\dot z + 2kQ - \frac{4}{729 \mathcal{B}^2} Q^3 - \rho_{m0}\right]dt.\end{equation}
The above action at once reveals that $z$ has been turned out to be cyclic and the symmetry may be obtained trivially from the field equations,

\begin{equation} \ddot Q = 0,\end{equation}

\begin{equation} \dot Q\dot z + 2k Q -\frac{4}{729 \mathcal{B}^2} Q^3 = \rho_{m0},\end{equation}
In view of the definition of the auxiliary variable $Q$  given in equation (45) and the field equation (47), one can find

\begin{equation} \frac{d}{dt} (a\sqrt R) = I,\end{equation}
where, $I$ is the same old conserved current obtained in (31). The conserved current so obtained in (31) or (49) may be solved at once to obtain

\begin{equation} \sqrt z = a = \left[a_{4}t^4 + a_{3} t^3 + \left(\frac{3 a_{3}^2}{8a_{4}} - k\right)t^2 + a_2 t + a_1\right]^{\frac{1}{2}},\end{equation}
This solution has been analyzed in detail in connection with our observable Universe and found to suffer from the disease that decelerated matter dominated epoch prior to the current accelerated epoch admits a solution $a \propto \sqrt t$, rather than the standard Friedmann solution $a \propto t^{\frac{2}{3}}$, which tells upon structure formation \cite{8}.\\

Summarily, Noether symmetry for $F(R)$ theory of gravity in Robertson-Walker space-time yields nothing other than $F(R) \propto R^{\frac{3}{2}}$ in vacuum or in the presence of matter in the form of dust along with a unique conserved current $I = \frac{d}{dt}(a\sqrt R)$. Despite this fact some authors \cite{12} have claimed to have obtained a number of conserved current in view of gauge Noether symmetry and keeping time translation in addition. Since Hamiltonian is the generator of time translation, so a generator in the form $\eta(t)\frac{\partial}{\partial t}$ does not lead to conserved Hamiltonian. On the contrary, a generator in the form $\frac{\partial}{\partial t}$, ie., $\eta = 1$, only leads to conserved Hamiltonian. However, in cosmology, the Hamiltonian is not only conserved but also is constrained to vanish and therefore is a part and parcel of the field equations. Thus, it is always required to set $\eta = 0$. Ignoring this very fact, the authors \cite{12} claim to have obtained a number of symmetry generators along with the corresponding conserved currents. Ridiculously, the authors \cite{12} do not take the trouble to check if such conserved currents satisfy the field equations. Earlier, some other authors have also made same type of mistakes and claimed to have obtained $F(R) \propto R^n$ ($n$ being arbitrary) in vacuum \cite{10}, and $F(R) \propto R^2$ along with a potential $V(\phi) \propto \phi^{-4}$ of a tachyonic field \cite{11}. Both the claims have been proved to be wrong \cite{13}, \cite{14}. We have emphasized that a non-zero gauge term does not improve the situation in any case rather than making things complicated. We have also shown that under the choice of an appropriate basic variable, viz., the first fundamental form $h_{ij} = a^2 = z$ and under the introduction of an auxiliary variable $Q = \frac{\partial S}{\partial \ddot z}$, the problem is oversimplified since $z$ becomes cyclic. The field equation, which is essentially the conserved momentum canonically conjugate to $h_{ij}$ and the only integral of motion of the theory is solved at once, exactly. The solution gives accelerated expansion of the Universe and fits SNIa data perfectly, nevertheless early deceleration tracks $a \propto \sqrt t$, rather than the standard Friedmann solution $a \propto t^{\frac{2}{3}}$. This result tells upon WMAP data at high redshift and also on the structure formation \cite{8}. Further, an action containing only $R^{\frac{3}{2}}$ term in the radiation dominated era results in a solution where, the scale factor tracks $a \propto t^{\frac{3}{4}}$ rather than the standard Friedmann solution $a \propto \sqrt t$ \cite{8}. Thus we conclude $R^{\frac{3}{2}}$ alone does not serve the purpose to explain presently available cosmological data. It has also been shown \cite{9} that the situation is improved under the addition of a linear (Einstein-Hilbert) term.

\end{document}